\batchmode
\makeatletter
\def\input@path{{"/home/jacob/Documents/Work/My Papers/Manifestly Covariant Lagrangians (2020)/"}}
\makeatother
\documentclass[twoside,twocolumn,english,prl,superscriptaddress,nobibnotes,nofootinbib]{revtex4-2}
\usepackage[latin9]{inputenc}
\usepackage{color}
\usepackage{babel}
\usepackage{mathrsfs}
\usepackage{mathtools}
\usepackage{amsmath}
\usepackage{amssymb}
\usepackage[pdftex,unicode=true,
 bookmarks=true,bookmarksnumbered=true,bookmarksopen=false,
 breaklinks=false,pdfborder={0 0 0},pdfborderstyle={},backref=false,colorlinks=true]
 {hyperref}
\hypersetup{pdftitle={Gauge Invariance for Classical Massless Particles with Spin},
 pdfauthor={Jacob A. Barandes},
 pdfsubject={Physics},
 pdfkeywords={Classical physics; particle physics; gauge theories; analytical dynamics},
 linkcolor=blue, urlcolor=blue, citecolor=blue}

\makeatletter
\usepackage{url}

\let\originalleft\left
\let\originalright\right
\renewcommand{\left}{\mathopen{}\mathclose\bgroup\originalleft}
\renewcommand{\right}{\aftergroup\egroup\originalright}

\def\smalloverbrace#1{\mathop{\vbox{\m@th\ialign{##\crcr%
      \noalign{\kern3\p@}%
      \tiny\downbracefill\crcr\noalign{\kern3\p@\nointerlineskip}%
      $\hfil\displaystyle{#1}\hfil$\crcr}}}\limits}

\def\smallunderbrace#1{\mathop{\vtop{\m@th\ialign{##\crcr
   $\hfil\displaystyle{#1}\hfil$\crcr
   \noalign{\kern3\p@\nointerlineskip}%
   \tiny\upbracefill\crcr\noalign{\kern3\p@}}}}\limits}

\makeatother

\begin{document}
\title{Gauge Invariance for Classical Massless Particles with Spin}
\author{Jacob A. Barandes}
\email{jacob\_barandes@harvard.edu}

\affiliation{Jefferson Physical Laboratory, Harvard University, Cambridge, MA 02138}
\date{\today}
\begin{abstract}
Wigner's quantum-mechanical classification of particle-types in terms
of irreducible representations of the Poincaré group has a classical
analogue, which we extend in this paper. We study the compactness
properties of the resulting phase spaces at fixed energy, and show
that in order for a classical massless particle to be physically sensible,
its phase space must feature a classical-particle counterpart of electromagnetic
gauge invariance. By examining the connection between massless and
massive particles in the massless limit, we also derive a classical-particle
version of the Higgs mechanism.
\end{abstract}
\maketitle

\global\long\def\vec#1{{\bf #1}}%
\global\long\def\vecgreek#1{\boldsymbol{#1}}%
\global\long\def\dotprod{\cdot}%
\global\long\def\crossprod{\times}%
\global\long\def\tud#1#2#3{{#1^{#2}}_{#3}}%
\global\long\def\tdu#1#2#3{{#1_{#2}}^{#3}}%
\global\long\def\defeq{\equiv}%
\global\long\def\Trace{\mathrm{Tr}}%
\global\long\def\transp{\mathrm{T}}%
\global\long\def\refvalue{0}%
\global\long\def\parens#1{(#1)}%
\global\long\def\bigparens#1{\big(#1\big)}%
\global\long\def\Bigparens#1{\Big(#1\Big)}%
\global\long\def\biggparens#1{\bigg(#1\bigg)}%
\global\long\def\Biggparens#1{\Bigg(#1\Bigg)}%
\global\long\def\bracks#1{[#1]}%
\global\long\def\bigbracks#1{\big[#1\big]}%
\global\long\def\Bigbracks#1{\Big[#1\Big]}%
\global\long\def\biggbracks#1{\bigg[#1\bigg]}%
\global\long\def\Biggbracks#1{\Bigg[#1\Bigg]}%
\global\long\def\curlies#1{\{#1\}}%
\global\long\def\bigcurlies#1{\big\{#1\big\}}%
\global\long\def\Bigcurlies#1{\Big\{#1\Big\}}%
\global\long\def\biggcurlies#1{\bigg\{#1\bigg\}}%
\global\long\def\Biggcurlies#1{\Bigg\{#1\Bigg\}}%
\global\long\def\verts#1{\vert#1\vert}%
\global\long\def\bigverts#1{\big\vert#1\big\vert}%
\global\long\def\Bigverts#1{\Big\vert#1\Big\vert}%
\global\long\def\biggverts#1{\bigg\vert#1\bigg\vert}%
\global\long\def\Biggverts#1{\Bigg\vert#1\Bigg\vert}%

\section{Introduction}

The ingredients of classical physics are usually simpler to visualize
and understand than those of quantum theory. Classical systems have
definite configurations that are related to physical properties like
spatial location, mass, momentum, and energy in a much more transparent
way than is the case for abstract quantum states. It is therefore
worthwhile to determine which seemingly quantum phenomena turn out
to have classical realizations, if only to clarify the underpinnings
of those phenomena without all the complexities that come along with
Hilbert spaces, and to help foster the kind of physical intuition
that can lead to new discoveries.

As an important example, intrinsic spin is often regarded as fundamentally
quantum in nature, but there exists a fully classical description
of relativistic point particles with arbitrary masses and fixed spin.
This classical description makes it possible to distinguish between
effects that are related to spin itself and effects that are connected
specifically to quantum mechanics.

With the eventual goal of explicating and extending this framework,\footnote{For a more comprehensive treatment of the results in this paper, see
\citep{Barandes:2019mcl}.} we begin in Section~\ref{sec:The-Manifestly-Covariant-Lagrangian-Formulation}
by suitably generalizing the usual Lagrangian formulation of classical
physics to a more expressly Lorentz-covariant form. In Section~\ref{sec:Transitive-Group-Actions-of-the-Poincare-Group},
we review the classification of particle-types in terms of transitive
group actions of the Poincaré group, expanding on earlier work \citep{BalachandranMarmoSkagerstamStern:1983gsfb,Souriau:1997sds,Rivas:2002ktsp}
and paralleling Wigner's classification \citep{Wigner:1939urilg}
of quantum particle-types in terms of irreducible Hilbert-space representations
of the Poincaré group. We will be most interested in the massless
case, for which we present new results that include the emergence
of a classical-particle form of electromagnetic gauge invariance.
In Section~\ref{sec:The-Massless-Limit}, we revisit this appearance
of gauge invariance from the perspective of the massive case in the
massless limit, along the way deriving a classical-particle version
of the Higgs mechanism, another novel result.

\section{The Manifestly Covariant Lagrangian Formulation\label{sec:The-Manifestly-Covariant-Lagrangian-Formulation}}

Consider a classical system with time parameter $t$, degrees of freedom
$q_{\alpha}$, Lagrangian $L$, and action functional 

\begin{equation}
S\bracks q\defeq\int dt\,L\parens{q,\dot{q},t}.\label{eq:ClassicalActionFromLagrangian}
\end{equation}
 Here dots denote derivatives with respect to the time $t$, and we
will assume that the system's configuration space is a linear manifold
with a global coordinate system given by the degrees of freedom $q_{\alpha}$.
Before we apply this framework to classical relativistic point particles,
we will find it useful to recast these ingredients in a form that
is more manifestly compatible with relativistic invariance.

To do so, we begin by replacing $t$ with an arbitrary smooth, monotonic
parameter $\lambda$. Letting dots now denote derivatives with respect
to $\lambda$, we can rewrite the action functional in the reparametrization-invariant
form\footnote{For an early example of this technique, see \citep{Dirac:1926rqmacs}.
For a more modern, pedagogical treatment, see \citep{DeriglazovRizzuti:2011rifcmse}.} 
\begin{equation}
S\bracks{q,t}\defeq\int d\lambda\,\mathscr{L}\parens{q,\dot{q},t,\dot{t}},\label{eq:ReparamInvActionFunctional}
\end{equation}
 where 
\begin{equation}
\mathscr{L}\parens{q,\dot{q},t,\dot{t}}\defeq\dot{t}\,L\parens{q,\dot{q}/\dot{t},t}.\label{eq:ReparamInvLagrangian}
\end{equation}
 Although we are not assuming that the degrees of freedom $q_{\alpha}$
have anything to do with physical space for now, it will be convenient
to introduce a Cartesian-like raised/lowered-index notation according
to 
\begin{equation}
\begin{aligned}q^{t} & \defeq c\,t, & q_{t} & \defeq-c\,t,\\
q^{\alpha} & \defeq q_{\alpha},\\
p^{t} & \defeq H/c, & p_{t} & \defeq-H/c,\\
p^{\alpha} & \defeq p_{\alpha},
\end{aligned}
\label{eq:DefRaisingLowerIndicesCanonicalVars}
\end{equation}
 where $p_{\alpha}$ are the system's usual canonical momenta, $H$
is the system's usual Hamiltonian derived from the original Lagrangian
$L$ in \eqref{eq:ClassicalActionFromLagrangian}, and $c$ is a constant
with units of energy divided by momentum. The quantities $p^{t}$
and $p^{\alpha}$ are then expressible in terms of the function \eqref{eq:ReparamInvLagrangian}
as 
\begin{equation}
p^{t}=\frac{\partial\mathscr{L}}{\partial\dot{q}_{t}},\quad p^{\alpha}=\frac{\partial\mathscr{L}}{\partial\dot{q}_{\alpha}},\label{eq:ReparamInvCanonicalMomenta}
\end{equation}
 and one can show that the Euler-Lagrange equations take the symmetric-looking
form 
\begin{equation}
\dot{p}^{t}=\frac{\partial\mathscr{L}}{\partial q_{t}},\quad\dot{p}^{\alpha}=\frac{\partial\mathscr{L}}{\partial q_{\alpha}}.\label{eq:ReparamInvParamDerivsCanonicalMomenta}
\end{equation}
 Moreover, the action functional \eqref{eq:ReparamInvActionFunctional}
now takes a form that resembles a Lorentz-covariant dot product involving
a square matrix $\eta\defeq\mathrm{diag}\parens{-1,1,\dotsc}$ that
naturally generalizes the Minkowski metric tensor in Cartesian coordinates
from special relativity, 
\begin{equation}
S\bracks q=\int d\lambda\,\bigparens{p_{t}\dot{q}^{t}+\sum_{\alpha}p_{\alpha}\dot{q}^{\alpha}}=\int d\lambda\,\begin{pmatrix}p^{t} & p^{\alpha}\end{pmatrix}\,\eta\,\begin{pmatrix}\dot{q}^{t}\\
\dot{q}^{\alpha}
\end{pmatrix},\label{eq:ReparamActionFunctionalFromCanonicalMomenta}
\end{equation}
 despite the fact that, again, the degrees of freedom $q_{\alpha}$
are not assumed at this point to have anything to do with physical
space. The action functional is then invariant under transformations
\begin{equation}
\begin{pmatrix}q^{t}\\
q^{\alpha}
\end{pmatrix}\mapsto\Lambda\begin{pmatrix}q^{t}\\
q^{\alpha}
\end{pmatrix},\quad\begin{pmatrix}p^{t}\\
p^{\alpha}
\end{pmatrix}\mapsto\Lambda\begin{pmatrix}p^{t}\\
p^{\alpha}
\end{pmatrix}\label{eq:ReparamGeneralizedLorentzTransform}
\end{equation}
 for square matrices $\Lambda$ satisfying the condition $\Lambda^{\transp}\eta\Lambda=\eta.$

Thus, this reparametrization-invariant Lagrangian formulation motivates
the introduction of phase-space variables $q^{t},q^{\alpha},p^{t},p^{\alpha}$
that transform covariantly under a generalized notion of Lorentz transformations.
We therefore refer to this framework as the manifestly covariant Lagrangian formulation
of our classical system's dynamics.

\section{Transitive Group Actions of the Poincaré Group\label{sec:Transitive-Group-Actions-of-the-Poincare-Group}}

Wigner showed in \citep{Wigner:1939urilg} that classifying the different
Hilbert spaces that provide irreducible representations of the Poincaré
group yields a systematic categorization of quantum-mechanical particle-types
into massive, massless, and tachyonic cases.\footnote{See \citep{Weinberg:1996tqtfi} for a pedagogical review.}
As shown in various treatments, such as \citep{BalachandranMarmoSkagerstamStern:1983gsfb,Souriau:1997sds,Rivas:2002ktsp},
there exists a classical analogue of this construction, one version
of which we review here. Toward the end of this section and in the
next section, we will present fundamental new results concerning previously
unexamined features of the massless case.

\subsection{Kinematics}

We start by laying out a formulation of the kinematics of a system
that we will eventually identify as a classical relativistic particle.

Given a classical system described by a manifestly covariant Lagrangian
formulation, we say that its phase space provides a transitive or
``irreducible'' group action of the Poincaré group (or serves as
a homogeneous space for the Poincaré group) if we can reach every
state $\parens{q,p}$ in the system's phase space by starting from
an arbitrary choice of reference state $\parens{q_{\refvalue},p_{\refvalue}}$
and acting with an appropriate Poincaré transformation $\parens{a,\Lambda}\in\mathbb{R}^{1,3}\rtimes O\parens{1,3}$,
where $a=\parens{a^{t},a^{x},a^{y},a^{z}}$ is a four-vector that
parametrizes translations in spacetime and $\Lambda$ is a Lorentz-transformation
matrix. The Poincaré group singles out systems whose phase spaces
consist of spacetime coordinates 

\begin{equation}
X^{\mu}\defeq\parens{c\,T,\vec X}^{\mu}\defeq\parens{c\,T,X,Y,Z}^{\mu}\label{eq:4DSpacetimeCoordinates}
\end{equation}
 and corresponding canonical four-momentum components 
\begin{equation}
p^{\mu}\defeq\frac{\partial\mathscr{L}}{\partial\dot{X}^{\mu}}\defeq\parens{E/c,\vec p}^{\mu},\label{eq:PoincareActionSystem4Momentum}
\end{equation}
 where we identify $H\defeq E$ as the system's energy. We will see
that such a system formalizes the notion of a classical relativistic
particle.

To be as general as possible, we allow the system to have an intrinsic
spin represented by an antisymmetric spin tensor $S$ with components
\begin{equation}
S^{\mu\nu}=-S^{\nu\mu},\label{eq:Def4DSpinTensor}
\end{equation}
 in terms of which we can define a proper three-vector $\tilde{\vec S}$
and a three-dimensional pseudovector $\vec S$ according to 
\begin{equation}
S^{\mu\nu}\defeq\begin{pmatrix}0 & \tilde{S}_{x} & \tilde{S}_{y} & \tilde{S}_{z}\\
-\tilde{S}_{x} & 0 & S_{z} & -S_{y}\\
-\tilde{S}_{y} & -S_{z} & 0 & S_{x}\\
-\tilde{S}_{z} & S_{y} & -S_{x} & 0
\end{pmatrix}^{\mathclap{\mu\nu}}.\label{eq:SpinTensorFrom3Vecs}
\end{equation}
 Hence, the system's phase space consists of states that we can denote
by $\parens{X,p,S}$ and that, by definition, behave under Poincaré
transformations $\parens{a,\Lambda}$ according to 

\begin{equation}
\parens{X,p,S}\mapsto\parens{\Lambda X+a,\Lambda p,\Lambda S\Lambda^{\transp}}.\label{eq:Def4DLorentzTransfState}
\end{equation}
 Taking our reference state to be 
\begin{equation}
\parens{0,p_{\refvalue},S_{\refvalue}}\label{eq:PhaseSpacePointRefState}
\end{equation}
 for convenient choices of $p_{\refvalue}^{\mu}$ and $S_{\refvalue}^{\mu\nu}$
that will be made on a case-by-case basis later, we can therefore
write each state of our system as 
\begin{equation}
\parens{X,p,S}\defeq\parens{a,\Lambda p_{\refvalue},\Lambda S_{\refvalue}\Lambda^{\transp}}.\label{eq:PhaseSpacePointFromPoincTransfFromRef}
\end{equation}
 Thus, the components $a^{\mu}$ of the four-vector $a$ and the entries
$\tud{\Lambda}{\mu}{\nu}$ of the Lorentz-transformation matrix $\Lambda$
effectively become the system's fundamental phase-space variables.

To keep our notation simple, we will refer to $a^{\mu}$ as $X^{\mu}$
in our work ahead, remembering that these variables are independent
of the Lorentz-transformation variables $\tud{\Lambda}{\mu}{\nu}$.
We will therefore express the functional dependence of the system's
manifestly covariant action functional as $S\bracks{X,\Lambda}$.

It is natural to introduce several derived tensors from the system's
fundamental physical quantities $X^{\mu},p^{\mu},S^{\mu\nu}$. We
define the system's orbital angular-momentum tensor $L$ by 
\begin{equation}
L^{\mu\nu}\defeq X^{\mu}p^{\nu}-X^{\nu}p^{\mu}=-L^{\nu\mu},\label{eq:Def4DOrbitalAngularMomentum}
\end{equation}
 and $L$ together with $S$ make up the system's total angular-momentum
tensor $J$: 
\begin{equation}
J^{\mu\nu}\defeq L^{\mu\nu}+S^{\mu\nu}=-J^{\nu\mu}.\label{eq:Def4DTotalAngMomTensor}
\end{equation}
 Defining the four-dimensional Levi-Civita symbol by 
\begin{align}
\epsilon_{\mu\nu\rho\sigma} & \defeq\begin{cases}
+1 & \textrm{for \ensuremath{\mu\nu\rho\sigma} an even permutation of \ensuremath{txyz}},\\
-1 & \textrm{for \ensuremath{\mu\nu\rho\sigma} an odd permutation of \ensuremath{txyz}},\\
0 & \textrm{otherwise}
\end{cases}\nonumber \\
 & =-\epsilon^{\mu\nu\rho\sigma},\label{eq:4DLeviCivita}
\end{align}
 the system's Pauli-Lubanski pseudovector $W$ is defined by 
\begin{equation}
W^{\mu}\defeq-\frac{1}{2}\epsilon^{\mu\nu\rho\sigma}p_{\nu}S_{\rho\sigma}=\parens{\vec p\dotprod\vec S,\ \parens{E/c}\vec S-\vec p\crossprod\tilde{\vec S}}^{\mu}.\label{eq:DefPauliLubanski4Vec}
\end{equation}
 The following quantities are then invariant under proper, orthochronous
Poincaré transformations, and therefore represent fixed features (or
Casimir invariants) of the system's phase space: 

\begin{align}
-m^{2}c^{2} & \defeq p_{\mu}p^{\mu},\label{eq:Def4DMassSquaredAsInvariant}\\
w^{2} & \defeq W_{\mu}W^{\mu},\label{eq:SquarePauliLubanskiAsInvariant}\\
s^{2} & \defeq\frac{1}{2}S_{\mu\nu}S^{\mu\nu}=\vec S^{2}-\tilde{\vec S}^{2},\label{eq:Def4DSpinSquaredAsInvariant}\\
\tilde{s}^{2} & \defeq\frac{1}{8}\epsilon_{\mu\nu\rho\sigma}S^{\mu\nu}S^{\rho\sigma}=\vec S\dotprod\tilde{\vec S}.\label{eq:Def4DDualSpinSquaredInvariant}
\end{align}

In the analogous quantum case, the third of these invariant quantities,
the spin-squared scalar $s^{2}$, would be quantized in increments
of $\hbar$ (or, more precisely, $\hbar^{2}$). In our classical context,
we are essentially working in the limit of large quantum numbers,
in which the correspondence principle holds and these quantities are
free to take on fixed values from a continuous set of real numbers.
Note, in particular, that the invariance of $s^{2}$ is entirely separate
from issues of quantization, just as the invariance of $m^{2}$ does
not require quantization.

\subsection{Dynamics}

We now turn to the system's dynamics.

In the absence of intrinsic spin, $S^{\mu\nu}=0$, the system's manifestly
covariant action functional is, from \eqref{eq:ReparamActionFunctionalFromCanonicalMomenta},
given by 
\begin{equation}
S_{\textrm{no\,spin}}\bracks{X,\Lambda}=\int d\lambda\,p_{\mu}\dot{X}^{\mu}=\int d\lambda\,\parens{\Lambda p_{0}}_{\mu}\dot{X}^{\mu}.\label{eq:ActionFunctionalNoSpin}
\end{equation}
 We will eventually need to establish a definite relationship between
the system's four-momentum $p^{\mu}$ and its four-velocity $\dot{X}^{\mu}\defeq dX^{\mu}/d\lambda$.

First, however, we will extend the action functional \eqref{eq:ActionFunctionalNoSpin}
to include intrinsic spin. We begin by introducing the standard Lorentz
generators: 
\begin{equation}
\tud{\bracks{\sigma_{\mu\nu}}}{\alpha}{\beta}=-i\delta_{\mu}^{\alpha}\eta_{\nu\beta}+i\eta_{\mu\beta}\delta_{\nu}^{\alpha}.\label{eq:LorentzGeneratorsMixedIndices}
\end{equation}
 Using the composition property of Lorentz transformations applied
to the case of infinitesimal shifts $\lambda\mapsto\lambda+d\lambda$
in the parameter $\lambda$, 
\begin{align}
\Lambda\parens{\lambda+d\lambda} & =\Lambda\parens{d\lambda}\Lambda\parens{\lambda}\nonumber \\
 & =\parens{1-\parens{i/2}d\theta^{\mu\nu}\parens{\lambda}\sigma_{\mu\nu}}\Lambda\parens{\lambda},\label{eq:InfinitesimalShiftLorentzTransf}
\end{align}
 where $d\theta^{\mu\nu}$ are the components of an antisymmetric
tensor of infinitesimal Lorentz boosts and angular displacements,
we have 
\begin{align}
\dot{\Lambda}\parens{\lambda} & \defeq\lim_{d\lambda\to0}\frac{\Lambda\parens{\lambda+d\lambda}-\Lambda\parens{\lambda}}{d\lambda}\nonumber \\
 & =-\frac{i}{2}\dot{\theta}^{\mu\nu}\parens{\lambda}\sigma_{\mu\nu}\Lambda\parens{\lambda}.\label{eq:ParamDerivLorentzTransf}
\end{align}
 Invoking the following trace identity satisfied by antisymmetric
tensors $A^{\mu\nu}=-A^{\nu\mu}$, 
\begin{equation}
\frac{1}{2}\Trace\bracks{\sigma^{\mu\nu}A}=iA^{\mu\nu},\label{eq:LorentzGeneratorsTracePeelsOffAntisymmTensor}
\end{equation}
or, more explicitly, 
\begin{equation}
\frac{1}{2}\tud{\bracks{\sigma^{\mu\nu}}}{\alpha}{\beta}\tud A{\beta}{\alpha}=iA^{\mu\nu},\label{eq:LorentzGeneratorsTracePeelsOffAntisymmTensorExplicit}
\end{equation}
 we can express the rates of change $\dot{\theta}^{\mu\nu}\parens{\lambda}$
according to 
\begin{equation}
\dot{\theta}^{\mu\nu}\parens{\lambda}=\frac{i}{2}\Trace\bracks{\sigma^{\mu\nu}\dot{\Lambda}\parens{\lambda}\Lambda^{-1}\parens{\lambda}}.\label{eq:ParamDerivFromTrace}
\end{equation}
 By an integration by parts, we can then recast the action functional
\eqref{eq:ActionFunctionalNoSpin} (up to an irrelevant boundary term)
as 
\begin{equation}
S_{\textrm{no spin}}\bracks{X,\Lambda}=\int d\lambda\,\frac{1}{2}L_{\mu\nu}\dot{\theta}^{\mu\nu}.\label{eq:ActionFunctionalNoSpinFromOrbAngMom}
\end{equation}

With the alternative form \eqref{eq:ActionFunctionalNoSpinFromOrbAngMom}
of the action functional in hand, we can straightforwardly introduce
intrinsic spin into the system's dynamics by making the replacement
$L_{\mu\nu}\mapsto J_{\mu\nu}\defeq L_{\mu\nu}+S_{\mu\nu}$. Converting
the term involving $L_{\mu\nu}$ back into the form \eqref{eq:ActionFunctionalNoSpin},
we thereby obtain the new action functional 
\begin{equation}
S\bracks{X,\Lambda}=\int d\lambda\,\mathscr{L}=\int d\lambda\,\biggparens{p_{\mu}\dot{X}^{\mu}+\frac{1}{2}\Trace\bracks{S\dot{\Lambda}\Lambda^{-1}}},\label{eq:ActionFunctionalWithSpin}
\end{equation}
 which now properly accounts for intrinsic spin. Explicitly, the last
term, which encodes the system's intrinsic spin, is given by 
\begin{equation}
\frac{1}{2}\Trace\bracks{S\dot{\Lambda}\Lambda^{-1}}\defeq\frac{1}{2}\tud S{\alpha}{\beta}\tud{\dot{\Lambda}}{\beta}{\gamma}\tud{\parens{\Lambda^{-1}}}{\gamma}{\alpha}.\label{eq:ActionTermSpinExplicit}
\end{equation}

The equations of motion derived from this action functional are 
\begin{align}
\dot{p}^{\mu} & =0,\label{eq:4MomEOM}\\
\dot{J}^{\mu\nu} & =0,\label{eq:SpinTensorEOM}
\end{align}
 and respectively express conservation of four-momentum and conservation
of total angular momentum, in keeping with Noether's theorem and the
symmetries of the dynamics under Poincaré transformations. It follows
that the Pauli-Lubanski pseudovector \eqref{eq:DefPauliLubanski4Vec}
is conserved, $\dot{W}^{\mu}=0$, and that the scalar quantities $-m^{2}c^{2}$
and $w^{2}$ defined in \eqref{eq:Def4DMassSquaredAsInvariant} and
\eqref{eq:SquarePauliLubanskiAsInvariant} are guaranteed to be constant,
as required.

As shown in \citep{SkagerstamStern:1981ldccps}, constancy of the
spin-squared scalar $s^{2}$ defined in \eqref{eq:Def4DSpinSquaredAsInvariant}
requires the imposition of an important Poincaré-invariant condition
on the system's phase space. To see why, we make use of the equation
of motion \eqref{eq:SpinTensorEOM} to compute the rate of change
of $s^{2}$: 
\[
\frac{d}{d\lambda}\biggparens{\frac{1}{2}S_{\mu\nu}S^{\mu\nu}}=S_{\mu\nu}\dot{S}^{\mu\nu}=2\dot{X}^{\nu}p^{\mu}S_{\mu\nu}=0.
\]
 Keep in mind that without a definite relationship between the four-momentum
$p^{\mu}$ and the four-velocity $\dot{X}^{\mu}$, this condition
is nontrivial. Because it establishes a constraint on all solution
trajectories in the system's phase space, we conclude that the following
Lorentz-invariant condition must hold: 

\begin{equation}
p_{\mu}S^{\mu\nu}=0.\label{eq:FourMomSpinTensorZeroPhysicalCondition}
\end{equation}
 Combined with the system's equations of motion \eqref{eq:4MomEOM}
and \eqref{eq:SpinTensorEOM}, this condition yields a pair of basic
relationships between the system's four-momentum $p^{\mu}$ and its
otherwise-unfixed four-velocity $\dot{X}^{\mu}$, 
\begin{align}
p\dotprod\dot{X} & =\pm mc^{2}\sqrt{-\dot{X}^{2}/c^{2}},\label{eq:FourMomDotProdFourVelRelationship}\\
m\sqrt{-\dot{X}^{2}/c^{2}}\,p^{\mu} & =\mp m^{2}\dot{X}^{\mu},\label{eq:FourMomFourVelSqrRootRelationship}
\end{align}
 where $p\dotprod\dot{X}\defeq p_{\mu}\dot{X}^{\mu}$ and $\dot{X}^{2}\defeq\dot{X}_{\mu}\dot{X}^{\mu}$.
The equations \eqref{eq:4MomEOM}\textendash \eqref{eq:FourMomFourVelSqrRootRelationship}
complete our specification of the system's dynamics.

Notice that the self-consistency condition \eqref{eq:FourMomSpinTensorZeroPhysicalCondition},
$p_{\mu}S^{\mu\nu}=0$, is phrased entirely in terms of ingredients
that have clear counterparts in classical field theory and in quantum
theory\textemdash namely, linear and angular momentum. As we will
see shortly, the condition \eqref{eq:FourMomSpinTensorZeroPhysicalCondition}
eliminates unphysical spin states that formally arise due to our use
of a manifestly Lorentz-covariant formalism, and thereby serves a
role that is closely related to the Lorenz equation $\partial_{\mu}A^{\mu}=0$
that appears both in the Proca field theory of a massive spin-1 boson
and as the Lorenz-gauge condition in electromagnetism.

Indeed, for a plane-wave mode of the form $A^{\mu}\parens x=\varepsilon^{\mu}\exp\parens{ip\dotprod x/\hbar}$
for a spin-1 field theory, where $\varepsilon^{\mu}$ is the wave's
polarization four-vector and encodes the wave's underlying spin,
the Lorenz equation reduces to $p_{\mu}\varepsilon^{\mu}=0$, thereby
eliminating one linear combination of polarizations and therefore
one independent spin state from the underlying spin-1 boson. By contrast,
our classical particle has a fixed but not quantized overall spin
\eqref{eq:Def4DSpinSquaredAsInvariant}, and the self-consistency
condition $p_{\mu}S^{\mu\nu}=0$ removes a continuous infinity of
unphysical spin states.

\subsection{Classification of the Transitive Group Actions}

Specializing to the \emph{orthochronous} Poincaré group, classifying
the different systems whose phase spaces give transitive group actions
is a straightforward exercise that parallels Wigner's approach in
\citep{Wigner:1939urilg}. As derived in detail in \citep{Barandes:2019mcl},
one finds that each such system can describe a massive particle $m^{2}>0$
or a massless particle $m^{2}=0$ with either positive energy $E=p^{t}c>0$
or negative energy $E=p^{t}c<0$, or a tachyon $m^{2}<0$, or the
vacuum $p^{\mu}=0$. Furthermore, the relations \eqref{eq:FourMomDotProdFourVelRelationship}
and \eqref{eq:FourMomFourVelSqrRootRelationship} imply that for each
of these cases, the four-momentum is parallel to the four-velocity,
$p^{\mu}\propto\dot{X}^{\mu}$. It then follows immediately from the
equations of motion \eqref{eq:4MomEOM} and \eqref{eq:SpinTensorEOM}
that $L^{\mu\nu}$ and $S^{\mu\nu}$ are separately conserved.

For a massive particle, we can take the reference state \eqref{eq:PhaseSpacePointRefState}
to describe the particle at rest, with reference four-momentum 
\begin{equation}
p_{\refvalue}^{\mu}=\parens{mc,\vec 0}^{\mu}.\label{eq:DefMassivePosEnergyRef4Mom}
\end{equation}
 The condition \eqref{eq:FourMomSpinTensorZeroPhysicalCondition}
then eliminates unphysical spin degrees of freedom and implies that
the particle's spin tensor \eqref{eq:SpinTensorFrom3Vecs} reduces
to the three-dimensional spin pseudovector $\vec S$, whose possible
orientations fill out a compact, fixed-energy region of the particle's
phase space.

By contrast, for massless particles and tachyons, the little group
of Lorentz transformations that preserve the particle's reference
four-momentum $p_{0}^{\mu}$ dictates that the particle's phase space
at any fixed energy is seemingly noncompact, leading to infinite entropies
and other thermodynamic pathologies, besides problems that arise in
the corresponding quantum field theories.\footnote{See, for example, \citep{Wigner:1963iqmeom,Abbott:1976mpcsi}, but
also \citep{SchusterToro:2013tcspwsfsa} for a more optimistic take.} For a tachyon, the only way to eliminate this noncompactness is to
require that the spin tensor vanishes, $S^{\mu\nu}=0$, meaning that
tachyons are naturally spinless.

For a massless particle, by contrast, the story is more interesting.
We can take the massless particle's reference four-momentum to be
\begin{equation}
p_{0}^{\mu}=\parens{E/c,0,0,E/c}^{\mu},\label{eq:DefMasslesPosEnergyRef4Mom}
\end{equation}
 and the phase-space self-consistency condition \eqref{eq:FourMomSpinTensorZeroPhysicalCondition},
$p_{\mu}S^{\mu\nu}=0$, then implies the corresponding reference spin
tensor 
\begin{equation}
S_{\refvalue}^{\mu\nu}=\begin{pmatrix}0 & S_{\refvalue,y} & -S_{\refvalue,x} & 0\\
-S_{\refvalue,y} & 0 & S_{\refvalue,z} & -S_{\refvalue,y}\\
S_{\refvalue,x} & -S_{\refvalue,z} & 0 & S_{\refvalue,x}\\
0 & S_{\refvalue,y} & -S_{\refvalue,x} & 0
\end{pmatrix}^{\mathclap{\mu\nu}}.\label{eq:MasslessPosEnergyRefSpinTensor}
\end{equation}
The most general little-group transformation preserving the reference
four-momentum \eqref{eq:DefMasslesPosEnergyRef4Mom} consists of a
Lorentz-transformation matrix $\Lambda$ of the form\footnote{For a derivation, see, for example, \citep{Barandes:2019mcl,Weinberg:1996tqtfi}.}
\begin{equation}
\Lambda\parens{\alpha,\beta,\theta}=L\parens{\alpha,\beta}R\parens{\theta},\label{eq:MasslessPosEnergyLittleGroupMatrixFactorized}
\end{equation}
 where 
\begin{equation}
R\parens{\theta}\defeq\begin{pmatrix}1 & 0 & 0 & 0\\
0 & \cos\theta & \sin\theta & 0\\
0 & -\sin\theta & \cos\theta & 0\\
0 & 0 & 0 & 1
\end{pmatrix}\label{eq:MasslessPosEnergyLittleGroupMatrixRotationZ}
\end{equation}
 is a pure rotation by an angle $\theta$ around the $z$ axis and
where 
\begin{equation}
L\parens{\alpha,\beta}\defeq\begin{pmatrix}1+\zeta & \alpha & \beta & -\zeta\\
\alpha & 1 & 0 & -\alpha\\
\beta & 0 & 1 & -\beta\\
\zeta & \alpha & \beta & 1-\zeta
\end{pmatrix}\label{eq:MasslessPosEnergyLittleGroupMatrixNoncompact}
\end{equation}
 is a complicated combination of Lorentz boosts and rotations. One
can show that 
\begin{align}
R\parens{\theta_{1}}R\parens{\theta_{2}} & =R\parens{\theta_{1}+\theta_{2}},\label{eq:MasslessPosEnergyRotationZCommutative}\\
L\parens{\alpha_{1},\beta_{1}}L\parens{\alpha_{2},\beta_{2}} & =L\parens{\alpha_{1}+\alpha_{2},\beta_{1}+\beta_{2}},\label{eq:MasslessPosEnergyNoncompactTransfCommutative}
\end{align}
 so rotations $R\parens{\theta}$ around the $z$ axis and the Lorentz
transformations $L\parens{\alpha,\beta}$ respectively form a pair
of commutative subgroups of the particle's little group. Noting that
\begin{align}
 & R\parens{\theta}L\parens{\alpha,\beta}R^{-1}\parens{\theta}\nonumber \\
 & \quad=L\parens{\alpha\cos\theta+\beta\sin\theta,-\alpha\sin\theta+\beta\cos\theta},\label{eq:MasslessPosEnergyLittleGroupNoncomm}
\end{align}
 we identify the little group as $ISO\parens 2$, which is the noncompact
group of rotations and translations in the two-dimensional Euclidean
plane. 

These little-group transformations act nontrivially on the particle's
reference spin tensor \eqref{eq:MasslessPosEnergyRefSpinTensor}:
\begin{align}
 & L\parens{\alpha,\beta}S_{\refvalue}L^{\transp}\parens{\alpha,\beta}\nonumber \\
 & =S_{\refvalue}+\begin{pmatrix}0 & -\beta S_{\refvalue,z} & \alpha S_{\refvalue,z} & 0\\
\beta S_{\refvalue,z} & 0 & 0 & \beta S_{\refvalue,z}\\
\alpha S_{\refvalue,z} & 0 & 0 & -\alpha S_{\refvalue,z}\\
0 & -\beta S_{\refvalue,z} & \alpha S_{\refvalue,z} & 0
\end{pmatrix}.\label{eq:MasslessPosEnergyNoncompactTransfSpinTensor}
\end{align}
 Hence, the only way to ensure that the massless particle has a compact
phase space at fixed reference energy while still allowing for nonzero
spin is to impose the following equivalence relation on the particle's
phase space: 
\begin{equation}
\parens{X,p,S}\cong\parens{X,p,S^{\prime}}.\label{eq:MasslessPosEnergyEquivRelationGaugeInvarianceGeneral}
\end{equation}

This equivalence relation is a new result. Just as the self-consistency
condition \eqref{eq:FourMomSpinTensorZeroPhysicalCondition}, $p_{\mu}S^{\mu\nu}=0$,
is the classical-particle analogue of the Lorenz-gauge condition $\partial_{\mu}A^{\mu}=0$
for the gauge potential $A_{\mu}$ in electromagnetism, the equivalence
relation \eqref{eq:MasslessPosEnergyEquivRelationGaugeInvarianceGeneral}
is a classical-particle manifestation of electromagnetic gauge invariance
$A_{\mu}\cong A_{\mu}+\partial_{\mu}f$. Indeed, for the case of plane
waves $A^{\mu}=\varepsilon^{\mu}\exp\parens{ip\dotprod x/\hbar}$
and $f=\alpha\exp\parens{ip\dotprod x/\hbar}$, where the polarization
four-vector $\varepsilon^{\mu}$ encodes the wave's underlying spin,
electromagnetic gauge invariance reduces to an equivalence relation
of the form $\varepsilon^{\mu}\cong\varepsilon^{\mu}+\parens{i\alpha/\hbar}p^{\mu}$
for the wave's polarization, and therefore implies an equivalence
relation on the wave's underlying spin directly analogous to \eqref{eq:MasslessPosEnergyEquivRelationGaugeInvarianceGeneral}.

In particular, in the same sense in which electromagnetic gauge invariance
is responsible for removing unphysical spin states, and furthermore
implies that all observable quantities must be gauge invariant, the
equivalence relation \eqref{eq:MasslessPosEnergyEquivRelationGaugeInvarianceGeneral}
cuts the classical massless particle's phase space at fixed energy
down to a compact extent, with the implication that \eqref{eq:MasslessPosEnergyEquivRelationGaugeInvarianceGeneral}
must be an invariance of all observable quantities. The distinct physical
states of the massless particle are then characterized by a spacetime
position $X^{\mu}$, a four-momentum $p^{\mu}$, and a helicity $\sigma\defeq\parens{\vec p/\verts{\vec p}}\dotprod\vec S$.\footnote{Note that if we permit parity transformations, which map $\sigma\mapsto-\sigma$,
then we must require that the equivalence relation \eqref{eq:MasslessPosEnergyEquivRelationGaugeInvarianceGeneral}
hold only for states that share the same helicity $\sigma$.} 

Components of the spin tensor $S^{\mu\nu}$ that are transverse to
the particle's three-momentum $\vec p$ are not invariant under the
equivalence relation \eqref{eq:MasslessPosEnergyEquivRelationGaugeInvarianceGeneral}.
As a consequence, $S^{\mu\nu}$ cannot directly appear in Lorentz-covariant
interaction terms in equations of motion that couple the particle
to other systems, in close analogy with the fact that gauge invariance
precludes the electromagnetic gauge potential $A_{\mu}$ from showing
up directly in Lorentz-covariant field equations outside of the gauge-invariant
Faraday tensor $F_{\mu\nu}\defeq\partial_{\mu}A_{\nu}-\partial_{\nu}A_{\mu}$.

\section{The Massless Limit\label{sec:The-Massless-Limit}}

We can better understand the origin of the novel equivalence relation
\eqref{eq:MasslessPosEnergyEquivRelationGaugeInvarianceGeneral} by
starting with the massive case $m>0$ and then taking the massless
limit $m\to0$.

Our original massive-particle reference state \eqref{eq:DefMassivePosEnergyRef4Mom}
degenerates for $m\to0$, so we instead take the massive particle's
reference four-momentum to be 
\begin{align}
\bar{p}^{\mu} & \defeq\parens{\bar{p}^{t},0,0,\bar{p}^{z}}^{\mu}=\bigparens{\sqrt{\parens{\bar{p}^{z}}^{2}+m^{2}c^{2}},0,0,\bar{p}^{z}}^{\mu}.\label{eq:DefMassivePosEnergyRef4MomAlt}
\end{align}
 This choice has the correct $m\to0$ limit \eqref{eq:DefMasslesPosEnergyRef4Mom}:
\begin{equation}
\lim_{m\to0}\bar{p}^{\mu}=\parens{E_{\refvalue}/c,0,0,E_{\refvalue}/c}^{\mu},\quad E_{\refvalue}\defeq\bar{p}^{z}c.\label{eq:MassivePosEnergyMasslessLimitFourMom}
\end{equation}
 Moreover, \eqref{eq:DefMassivePosEnergyRef4MomAlt} is related to
our original choice \eqref{eq:DefMassivePosEnergyRef4Mom} of reference
four-momentum for the massive particle by a simple Lorentz boost $\bar{\Lambda}$
along the $z$ direction, 
\begin{equation}
\bar{p}^{\mu}=\tud{\bar{\Lambda}}{\mu}{\nu}p_{\refvalue}^{\nu},\label{eq:MassivePosEnergyFourMomAltFromLorentzBoost}
\end{equation}
 and the new reference value $\bar{S}$ of the massive particle's
spin tensor is related to its original reference value $S_{\refvalue}$
according to 
\begin{align}
 & \bar{S}^{\mu\nu}\defeq\parens{\bar{\Lambda}S_{\refvalue}\bar{\Lambda}^{\transp}}^{\mu\nu}\nonumber \\
 & =\begin{pmatrix}0 & {\displaystyle \frac{\bar{p}^{z}}{mc}}S_{\refvalue,y} & -{\displaystyle \frac{\bar{p}^{z}}{mc}}S_{\refvalue,x} & 0\\
-{\displaystyle \frac{\bar{p}^{z}}{mc}}S_{\refvalue,y} & 0 & S_{\refvalue,z} & -{\displaystyle \frac{\bar{p}^{t}}{mc}}S_{\refvalue,y}\\
{\displaystyle \frac{\bar{p}^{z}}{mc}}S_{\refvalue,x} & -S_{\refvalue,z} & 0 & {\displaystyle \frac{\bar{p}^{t}}{mc}}S_{\refvalue,x}\\
0 & {\displaystyle \frac{\bar{p}^{t}}{mc}}S_{\refvalue,y} & -{\displaystyle \frac{\bar{p}^{t}}{mc}}S_{\refvalue,x} & 0
\end{pmatrix}^{\mathclap{\mu\nu}}.\label{eq:MassivePosEnergyNewRefSpinTensorFromBoostOld}
\end{align}

For $m\to0$, we have $\bar{p}^{t},\bar{p}^{z}\to E_{\refvalue}/c$,
so the components $\bar{S}^{\mu\nu}$ of the spin tensor involving
$\bar{p}^{t}/mc$ or $\bar{p}^{z}/mc$ diverge. Furthermore, there
is a discrete mismatch in the particle's spin-squared scalar \eqref{eq:Def4DSpinSquaredAsInvariant}
between the massive case and the massless case: 
\begin{align}
s^{2}=S_{\refvalue,x}^{2}+S_{\refvalue,y}^{2} & +S_{\refvalue,z}^{2}\quad\parens{\textrm{massive}}\nonumber \\
 & \ne S_{\refvalue,z}^{2}\quad\parens{\textrm{massless}}.\label{eq:MassivePosEnergySpinSqDisagreeWithMassless}
\end{align}
 These discrepancies are hints that the massive case includes spin
degrees of freedom that need to be removed before taking the massless
limit.

Our approach for removing these ill-behaved spin degrees of freedom
is motivated by a corresponding procedure in quantum field theory
that was originally developed by Stueckelberg in \citep{Stueckelberg:1938dwefk}.
We start with the redefinition 
\begin{equation}
\begin{pmatrix}\bar{S}_{x}\\
\bar{S}_{y}
\end{pmatrix}\mapsto\frac{mc}{\bar{p}^{t}}\begin{pmatrix}\bar{S}_{x}+\bar{p}^{t}\varphi_{x}\\
\bar{S}_{y}+\bar{p}^{t}\varphi_{y}
\end{pmatrix}=\frac{mc}{\bar{p}^{t}}\begin{pmatrix}\bar{S}_{x}\\
\bar{S}_{y}
\end{pmatrix}+mc\begin{pmatrix}\varphi_{x}\\
\varphi_{y}
\end{pmatrix},\label{eq:MassivePosEnergyStueckelbergRedefSpinTensor}
\end{equation}
 where $\varphi_{x}\parens{\lambda}$ and $\varphi_{y}\parens{\lambda}$
are arbitrary new functions on the particle's worldline. The particle's
spin tensor \eqref{eq:MassivePosEnergyNewRefSpinTensorFromBoostOld}
then has the decomposition 
\begin{align}
\bar{S}^{\mu\nu} & =\begin{pmatrix}0 & {\displaystyle \frac{\bar{p}^{z}}{\bar{p}^{t}}}S_{\refvalue,y} & -{\displaystyle \frac{\bar{p}^{z}}{\bar{p}^{t}}}S_{\refvalue,x} & 0\\
-{\displaystyle \frac{\bar{p}^{z}}{\bar{p}^{t}}}S_{\refvalue,y} & 0 & S_{\refvalue,z} & -S_{\refvalue,y}\\
{\displaystyle \frac{\bar{p}^{z}}{\bar{p}^{t}}}S_{\refvalue,x} & -S_{\refvalue,z} & 0 & S_{\refvalue,x}\\
0 & S_{\refvalue,y} & -S_{\refvalue,x} & 0
\end{pmatrix}^{\mathclap{\mu\nu}}\nonumber \\
 & +\begin{pmatrix}0 & \bar{p}^{z}\varphi_{y} & -\bar{p}^{z}\varphi_{x} & 0\\
-\bar{p}^{z}\varphi_{y} & 0 & 0 & -\bar{p}^{t}\varphi_{y}\\
\bar{p}^{z}\varphi_{x} & 0 & 0 & \bar{p}^{t}\varphi_{x}\\
0 & \bar{p}^{t}\varphi_{y} & -\bar{p}^{t}\varphi_{x} & 0
\end{pmatrix}^{\mathclap{\mu\nu}},\label{eq:MassivePosEnergyRedefinedSpinTensor}
\end{align}
 and the spin-squared scalar \eqref{eq:Def4DSpinSquaredAsInvariant}
becomes 
\begin{align}
s^{2} & =\Biggparens{1-\Biggparens{\frac{\bar{p}^{z}}{\bar{p}^{t}}}^{2}}\Big(\parens{S_{\refvalue,x}+\bar{p}^{t}\varphi_{x}}^{2}\nonumber \\
 & \qquad\qquad\qquad\qquad+\parens{S_{\refvalue,y}+\bar{p}^{t}\varphi_{y}}^{2}\Big)+S_{\refvalue,z}^{2}.\label{eq:MassivePosEnergyRedefinedSpinSqScalar}
\end{align}
 The particle's spin tensor \eqref{eq:MassivePosEnergyRedefinedSpinTensor}
is now invariant under the simultaneous transformations 
\begin{align}
\begin{pmatrix}\bar{S}_{x}\\
\bar{S}_{y}
\end{pmatrix} & \mapsto\begin{pmatrix}\bar{S}_{x}\\
\bar{S}_{y}
\end{pmatrix}-\bar{p}^{t}\begin{pmatrix}f_{x}\\
f_{y}
\end{pmatrix},\label{eq:MassivePosEnergyGaugeTransfPerpSpin}\\
\begin{pmatrix}\varphi_{x}\\
\varphi_{y}
\end{pmatrix} & \mapsto\begin{pmatrix}\varphi_{x}\\
\varphi_{y}
\end{pmatrix}+\begin{pmatrix}f_{x}\\
f_{y}
\end{pmatrix},\label{eq:MassivePosEnergyGaugeTransfAncillary}
\end{align}
 where $f_{x}\parens{\lambda},f_{y}\parens{\lambda}$ are arbitrary
functions on the particle's worldline.

Our massive particle's original phase space, with states labeled as
$\parens{X,p,S}$, is therefore equivalent to a \emph{formally enlarged}
phase space consisting of states $\parens{X,p,S,\varphi}$ under the
equivalence relation $\parens{\bar{X},\bar{p},\bar{S},\varphi}\cong\parens{\bar{X},\bar{p},\bar{S}-\bar{p}^{t}f,\varphi+f}$,
suitably generalized from the reference state $\parens{\bar{X},\bar{p},\bar{S},\varphi}$
to general states $\parens{X,p,S,\varphi}$ of the system. Indeed,
one can check that the specific choice $\parens{f_{x},f_{y}}\defeq-\parens{\varphi_{x},\varphi_{y}}$
yields $\parens{\bar{X},\bar{p},\bar{S}+\bar{p}^{t}\varphi,0}$, which
gives back the state $\parens{\bar{X},\bar{p},\bar{S}}$ after undoing
the redefinition \eqref{eq:MassivePosEnergyStueckelbergRedefSpinTensor}
of $\bar{S}^{\mu\nu}$.

We can now safely take the massless limit of the system's redefined
spin tensor \eqref{eq:MassivePosEnergyRedefinedSpinTensor}: 
\begin{align}
\lim_{m\to0}\bar{S}^{\mu\nu} & =\begin{pmatrix}0 & S_{\refvalue,y} & -S_{\refvalue,x} & 0\\
-S_{\refvalue,y} & 0 & S_{\refvalue,z} & -S_{\refvalue,y}\\
S_{\refvalue,x} & -S_{\refvalue,z} & 0 & S_{\refvalue,x}\\
0 & S_{\refvalue,y} & -S_{\refvalue,x} & 0
\end{pmatrix}^{\mathclap{\mu\nu}}\nonumber \\
 & +\frac{E}{c}\begin{pmatrix}0 & \varphi_{y} & -\varphi_{x} & 0\\
-\varphi_{y} & 0 & 0 & -\varphi_{y}\\
\varphi_{x} & 0 & 0 & \varphi_{x}\\
0 & \varphi_{y} & -\varphi_{x} & 0
\end{pmatrix}^{\mathclap{\mu\nu}}\label{eq:MassivePosEnergyRedefinedSpinTensorMasslessLimit}
\end{align}
 and 
\begin{equation}
\lim_{m\to0}s^{2}=S_{\refvalue,z}^{2}.\label{eq:MassivePosEnergyRedefinedSpinSqMasslessLimit}
\end{equation}
 The degrees of freedom describing spin components perpendicular to
the particle's reference three-momentum $\bar{\vec p}$ no longer
contribute to the particle's spin-squared scalar $s^{2}$. If we remove
these ancillary degrees of freedom by setting $\varphi_{x},\varphi_{y}$
equal to zero, then the particle's spin tensor \eqref{eq:MassivePosEnergyRedefinedSpinTensorMasslessLimit}
reduces correctly to the reference spin tensor \eqref{eq:MasslessPosEnergyRefSpinTensor}
for a massless particle, and our equivalence relation \eqref{eq:MassivePosEnergyGaugeTransfPerpSpin}
reduces to the gauge invariance \eqref{eq:MasslessPosEnergyEquivRelationGaugeInvarianceGeneral}.
We have therefore completed our recovery of the massless case from
the $m\to0$ limit of a massive particle.

We can run these arguments in reverse to obtain a classical-particle
counterpart of the celebrated Higgs mechanism. Recall that in the
simplest version of the field-theoretic Higgs mechanism, we start
with a massless spin-1 boson and then spontaneously break the underlying
gauge invariance. The overall effect is to give the original spin-1
boson a nonzero mass while augmenting its two spin states with another
spin state from a Higgs boson, so that we now have the necessary three
spin states for a massive spin-1 boson. Analogously, suppose that
we start with a classical massless particle with reference four-momentum
\eqref{eq:DefMasslesPosEnergyRef4Mom}, $p_{0}^{\mu}=\parens{E/c,0,0,E/c}^{\mu}$,
and phase-space equivalence relation \eqref{eq:MasslessPosEnergyEquivRelationGaugeInvarianceGeneral},
$\parens{X,p,S}\cong\parens{X,p,S^{\prime}}$. Running the analysis
of this section in the other direction, we see that we can transform
the massless particle into a massive particle if we augment the particle's
phase space with ancillary ``Higgs'' degrees of freedom $\varphi_{x},\varphi_{y}$.

\section*{Acknowledgments}

J.\,A.\,B. has benefited from personal communications with Howard
Georgi, Andrew Strominger, David Griffiths, David Kagan, David Morin,
Logan McCarty, Monica Pate, and Alex Lupsasca.

\bibliographystyle{2_home_jacob_Documents_Work_My_Papers_Manifestl___ant_Lagrangians__2020__custom-abbrvunsrturl}
\bibliography{0_home_jacob_Documents_Work_My_Papers_Bibliography_Global-Bibliography}

\end{document}